\documentclass[twocolumn,showkeys,showpacs,amsmath,amssymb,superscriptaddress,preprintnumbers,nofootinbib]{revtex4}
\usepackage{amssymb}
\usepackage{amsmath}
\usepackage{graphicx}

\normalbaselines
\begin{document}

\title{The Einstein-Maxwell-aether-axion theory: \\ Dynamo-optical anomaly in the electromagnetic response }

\author{Timur Yu. Alpin}
\email{Timur.Alpin@kpfu.ru}
\author{Alexander B. Balakin}
\email{Alexander.Balakin@kpfu.ru}
\affiliation{Department of General Relativity and
Gravitation, Institute of Physics,Kazan Federal University, Kremlevskaya street 18, 420008, Kazan,
Russia}

\begin{abstract}
\noindent We consider a pp-wave symmetric model in the framework
of the Einstein-Maxwell-aether-axion theory. Exact solutions to
the equations of axion electrodynamics are obtained for the model,
in which pseudoscalar, electric and magnetic fields were constant
before the arrival of a gravitational pp-wave. We show that
dynamo-optical interactions, i.e., couplings of electromagnetic
field to a  dynamic unit vector field, attributed to the velocity
of a cosmic substratum (aether, vacuum, dark fluid...), provide
the  response of axionically active electrodynamic system to
display anomalous behavior.
\end{abstract}

\pacs{04.20.-q, 04.40.-b, 04.40.Nr, 04.50.Kd}

\keywords{unit vector field, dark matter axions, dynamo-optical phenomena}

\maketitle

\section{Introduction}

The term {\it dynamo-optical} phenomena was introduced in
\cite{LL} in order to distinguish specific electromagnetic effects
in media, which move non-uniformly (the macroscopic velocity field
of such media is characterized by local acceleration, shear,
vorticity and expansion).  Irregularities of motion of
electromagnetically active media can produce specific effects
analogous to classical birefringence, optical activity, etc.,
(see, e.g., \cite{Eringen,Hehlbook}). A natural question arises:
whether similar dynamo-optical phenomena can be displayed in {\it
cosmic} electrodynamic systems, e.g., in pulsar magnetospheres or
in super-critical black hole accretion disks? In other words,
whether one has a chance to find fingerprints of cosmic dynamics
in the properties of incoming light, which the astronomers are
studying? The discussion of this problem is faced with a
philosophic question: what is the origin of velocity field, the
irregularities of which one needs to study? It might be, for
instance, the velocity field, attributed to a supermassive black
hole, the velocity field of a baryonic matter, of a dark matter,
of a dark fluid, etc. The corresponding formalism is based on the
analysis of the time-like unit eigen four-vector of the
corresponding stress-energy tensor (see, e.g.,
\cite{BDehnen,BDolbilova,BAlpin}). This vector field is obtained
algebraically and thus requires additional variation procedures
(it was elaborated in \cite{B1,B2,BDolbilova}). An alternative
approach for studying of the dynamo-optical phenomena is connected
with the Einstein-Maxwell-aether theory established in
\cite{BL2014} on the base of Einstein-aether theory
\cite{J1,J2,J3,J4,J5,J6,J7,J8}. In these theories there is a
non-vanishing everywhere dynamic time-like unit vector field $U^i$
characterizing the velocity of a cosmic substratum (the vacuum,
the aether, the dark fluid and so on). The corresponding variation
procedure is standard for the relativistic field theory. The
Einstein-aether and the Einstein-Maxwell-aether theories realize
the idea of a preferred frame of reference \cite{CW,N1,N2}
associated with a world-line congruence for which the
corresponding time-like velocity four-vector $U^i$ is the tangent
vector. In this sense they are characterized by a violation of
Lorentz invariance (see, e.g., \cite{LIV31,O}). Based on this
approach one can reformulate the idea of appearance of the
dynamo-optical phenomena;  now they characterize a specific type
of interactions between the electromagnetic field and the unit
dynamic vector field. The main detail of the mathematical
description of this interaction is that the covariant derivative
of the unit vector field, $\nabla_i U_k$, enters the Lagrangian
linearly and in quadratic form.

In the paper \cite{BL2014} we mentioned three interesting
applications of the established Einstein-Maxwell-aether theory;
one of them is the application to the space-times with the
so-called pp-wave symmetry. Now we consider this application in
more details, using the restricted model. Generally, the full
model can not satisfy the requirements of the pp-wave symmetry.
Nevertheless, the Einstein-aether theory contains four Jacobson's
parameters, which are not yet determined and thus can be
considered as arbitrary, $C_1$, $C_2$, $C_3$ and $C_4$. If we use
two constraints, say, $C_2{=}0$ and $C_1{+}C_3{=}0$, the model, as
we have shown, can accept the pp-wave symmetry. Physically, this
means that the aether is considered as a substratum insensitive
with respect to a shear and expansion of the velocity field, but
remains sensitive to acceleration and vorticity. In this case the
analysis of the model is simplified very seriously, and we used
this model as an example of application. In order to explain the
interest to such model, we attract the attention to the results
obtained in \cite{BAlpin}, namely, that due to the dynamo-optical
interactions the electric and magnetic response of the system to
the action of the gravitation radiation can be anomalously strong.
The model under consideration, as it will be shown below, also
displays such behavior.

And the last new detail. We introduced into the
Einstein-Maxwell-aether theory a pseudoscalar field \cite{WTN77},
which could be (in principle) associated with axions and thus with
the cosmic dark matter (see, e.g., \cite{A1}- \cite{A16} for
details and references). Now the theory can be indicated as
Einstein-Maxwell-aether-axion theory. This pseudoscalar field
plays the role of mediator in the interaction between
gravitational-wave, electric, magnetic and unit dynamic vector
fields, providing the anomalous growth of the electromagnetic
response in analogy with the effects described in \cite{BWTN}. Let
us emphasize that in \cite{BWTN} the exact solutions to the
equations of the axion electrodynamics were obtained for the case,
when the background gravity field has the pp-wave symmetry, but
the effects of non-uniform motion of the medium were not
considered. From this point of view the presented work is the
generalization of the work \cite{BWTN}, which takes into account
the dynamo-optical interactions of the electrodynamic system with
the unit dynamic vector field.

The paper is organized as follows. In Section II we consider the
basic elements of the Einstein-aether theory supplemented by the
pseudoscalar field, and describe the background state possessing
the pp-wave symmetry. In Section III we extend the model by adding
the test electromagnetic field to the background field. In Section
IV we obtain exact solutions for the electric, magnetic and
pseudoscalar field. In Section V we discuss the anomalies in the
electromagnetic response of the system on the action of the
gravitational-pseudoscalar pp-waves. Conclusions are presented in
Section VI.

\section{Background state}\subsection{Action functional}

We consider the Einstein-Maxwell-aether-axion model in the
framework of a hierarchical approach. In our scheme three
constituents form the background state of the global physical
system: first, the gravity field; second, the unit dynamic vector
field attributed to the velocity of the aether; third, the
pseudoscalar (axion) field. The electromagnetic field is treated
to be the test subsystem, which is influenced by the background,
but does not change its state. In order to describe the background
state we use the action functional
$$
S_{({\rm B})} = \int d^4 x \sqrt{{-}g} \ \left\{
\frac{1}{2\kappa} \left[R{+} \lambda \left(g_{mn}U^m
U^n {-}1 \right) {+} \right. \right.
$$
$$
\left. \left. {+}K^{abmn}  \ \nabla_a U_m  \ \nabla_b U_n
\right]  {-} \frac{1}{2}  \xi \ \Psi^2_0 \ g^{mn} \ \nabla_m \phi \nabla_n \phi
{+} \right.
$$
\begin{equation}
\left. + \frac14 \Psi^2_0 \left[\frac{m^2_{({\rm a})}}{\nu} + \nu (\phi^2-\phi^2_{*})\right]^2  \right\} \,,
\label{1}
\end{equation}
where $g = {\rm det}(g_{ik})$ is the determinant of the metric,
$R$ is the Ricci scalar, and $\kappa$ is the Einstein constant.
Two last terms in this integral present the Lagrangian of a
pseudoscalar (axion) field. The pseudoscalar $\phi$ is considered
as a dimensionless quantity; the multiplier $\Psi_0$ brings the
dimension to the pseudoscalar field. The parameter $\xi$ takes two
values. First, when $\xi=1$, we deal with a pseudoscalar field
with the fourth order potential
\begin{equation}
V(\phi^2) = \frac14 \Psi^2_0 \left[\frac{m^2_{({\rm a})}}{\nu} + \nu (\phi^2-\phi^2_{*})\right]^2  \,.
\label{b1}
\end{equation}
This potential possesses the following two properties:
\begin{equation}
V(\Phi_{}^2) = 0 \,, \quad
\left[\frac{d}{d\phi}V(\phi_{}^2)\right]_{|\phi=\Phi} = 0 \,,
\label{b11}
\end{equation}
where the constant $\Phi$ takes one of the two values
\begin{equation}
\Phi = \pm \sqrt{\phi^2_{*} - \frac{m^2_{({\rm a})}}{\nu^2}} \,.
\label{b19}
\end{equation}
When $|\phi_{*}| > \frac{m_{({\rm a})}}{|\nu|}$, this potential
has two symmetric minima, each of them coincides with the corresponding double zero of the
function $V(\phi^2)$. Second, when $\xi=-1$, the pseudoscalar field is phantom
- like, or in other words, the field with negative kinetic term.
Other terms in the action functional (\ref{1}) involve the vector
field $U^i$. The first term of this type, $\lambda \left(g_{mn}U^m
U^n {-}1 \right)$, ensures that the $U^i$ is normalized to one.
The second term $K^{abmn} \ \nabla_a U_m \ \nabla_b U_n $ is
quadratic in the covariant derivative $\nabla_a U_m $ of the
vector field $U^i$, with $K^{abmn}$ a tensor field constructed
using the metric tensor $g^{ij}$ and the velocity four-vector
$U^k$ only \cite{J1},
$$
K^{abmn} =
$$
\begin{equation}
C_1 g^{ab} g^{mn} {+} C_2 g^{am}g^{bn}
{+} C_3 g^{an}g^{bm} {+} C_4 U^{a} U^{b}g^{mn} \,.
\label{2}
\end{equation}
The Jacobson constants $C_1$, $C_2$, $C_3$ and $C_4$ are
phenomenologically introduced \cite{J1,J2,J3}; there is a number
of proposals to estimate these constants from observations (see,
e.g., \cite{J4,J5}).

\subsection{Master equations describing the background state}

\subsubsection{Equations of aether dynamics}

The aether dynamic equations can be found by variation of the action functional (\ref{1}) with
respect to the Lagrange multiplier $\lambda$ and the vector field $U^i$.
The variation with respect to $\lambda$ yields the equation
\begin{equation}
g_{mn}U^m U^n = 1 \,,
\label{29}
\end{equation}
which is the normalization condition of the time-like vector field $U^k$.
The variation of the functional (\ref{1}) with respect to
$U^i$ yields the dynamic equation
\begin{equation}
\nabla_m {\cal J}^{mn}
- I^n  = \lambda \ U^n  \,.
\label{0A1}
\end{equation}
Here we are using the standard definitions \cite{J1}
\begin{equation}
{\cal J}^{mn} =  K^{lmsn} \nabla_l U_s \,,
\label{0A31}
\end{equation}
\begin{equation}
I^n =  \frac12 (\nabla_l U_s) (\nabla_m U_j) \
 \frac{\delta K^{lsmj}}{\delta U_n}  =  C_4  DU_m \nabla^n U^m
\,.
\label{0A4A}
\end{equation}
The operator $D$ appearing in (\ref{0A4A}) is the convective
derivative defined as $ D\equiv U^i\,\nabla_i$. The Lagrange
multiplier can be obtained by convolution of (\ref{0A1}) with
$U_n$ with normalization condition (\ref{29}):
\begin{equation}
\lambda = U_n \left[\nabla_m {\cal J}^{mn}
- I^n \right]  \,.  \label{0A309}
\end{equation}

\subsubsection{Equation for the pseudoscalar field}

The variation of the action functional (\ref{1}) with respect to pseudoscalar field $\phi$ gives the equation
\begin{equation}
\xi \nabla^k \nabla_k \phi +  \left[m^2_{({\rm a})} + \nu^2 (\phi^2-\phi^2_{*})\right] \phi = 0
\,. \label{axioneq}
\end{equation}
Clearly, the constant solution $\phi = \Phi$ satisfies this equation.  When $\nu=0$ and $\xi=1$, the equation (\ref{axioneq}) 
is the standard Klein-Gordon equation with the mass $m_{({\rm a})}$.

\subsubsection{Gravity field equations}

The variation of the action (\ref{1}) with respect to the metric
$g^{ik}$ yields the gravitational field equations
\begin{equation}
R_{ik} - \frac{1}{2} R \ g_{ik} = T^{({\rm U})}_{ik} +
\kappa T^{({\rm A})}_{ik} \,. \label{0Ein1}
\end{equation}
The term $T^{({\rm U})}_{ik}$ describes
the stress-energy tensor associated with the self-gravitation
of the vector field $U^i$; it has the form:
$$
T^{({\rm U})}_{ik} = C_1\left(\nabla_mU_i \nabla^m U_k {-}
\nabla_i U_m \nabla_k U^m \right) {+}
$$
$$
{+}C_4 DU_i DU_k  {+} U_iU_k
U_n \left[\nabla_m {\cal J}^{mn} {-} I^n
\right] {+}
$$
$$
{+}\nabla^m \left[U_{(i}{\cal J}_{k)m}  {-}
{\cal J}_{m(i} U_{k)}  {-}
{\cal J}_{(ik)} U_m\right] {+}
$$
\begin{equation}
{+}\frac12 g_{ik} {\cal J}^{am} \nabla_a U_m \,, \label{0Ein5}
\end{equation}
where $p_{(i} q_{k)}{\equiv}\frac12 (p_iq_k{+}p_kq_i)$
denotes symmetrization.
The term
$$
T^{({\rm A})}_{ik} =\Psi^2_0 \xi \nabla_i \phi \nabla_k \phi -
$$
\begin{equation}
{-} \frac{\Psi^2_0}{2} g_{ik} \left[\xi \nabla^m \phi \nabla_m
\phi {-} \frac12  \left[\frac{m^2_{({\rm a})}}{\nu} {+} \nu
(\phi^2{-}\phi^2_{*})\right]^2  \right] \label{Ta}
\end{equation}
describes the  stress-energy tensor of the pseudoscalar (axion)
field. As usual, the total stress-energy tensor satisfies the
relationship
\begin{equation}
\nabla^k\left[ T^{({\rm U})}_{ik} +
\kappa T^{({\rm A})}_{ik} \right] = 0
\,, \label{compa1}
\end{equation}
which is the identity on the solutions of (\ref{0A1}) and (\ref{axioneq}).

\subsection{Decomposition of the tensor $\Theta_{ik} \equiv \nabla_i U_k$}

The tensor $\Theta_{ik} \equiv \nabla_i U_k$ can be decomposed,
as usual, into a sum of its irreducible parts,
namely, the acceleration four-vector $DU^{i}$,
the symmetric trace-free shear tensor $\sigma_{ik}$,
the anti-symmetric vorticity tensor $\omega_{ik}$, and
the expansion scalar $\Theta=\Theta^k_{\ k}$. The decomposition is given by
\begin{equation}
\Theta_{ik} = \nabla_i U_k = U_i DU_k + \sigma_{ik} + \omega_{ik} +
\frac{1}{3} \Delta_{ik} \Theta \,, \label{act3}
\end{equation}
where the basic quantities are defined as follows:
$$
DU_k \equiv U^m \nabla_m U_k \,,  \Theta \equiv \nabla_m U^m  \,,  \Delta^i_k = \delta^i_k {-}U^i U_k \,,
$$
$$
\sigma_{ik}
\equiv \frac{1}{2}\Delta_i^m \Delta_k^n \left(\nabla_m U_n {+}
\nabla_n U_m \right) {-} \frac{1}{3}\Delta_{ik} \Theta  \,,
$$
\begin{equation}
\omega_{ik} \equiv \frac{1}{2}\Delta_i^m \Delta_k^n \left(\nabla_m
U_n {-} \nabla_n U_m \right) \,. \label{act4}
\end{equation}
In these terms the dynamic tensor ${\cal J}^{mn}$ takes the form:
$$
{\cal J}^{mn} = (C_1 {+} C_4)U^m DU^n {+} C_3 U^n DU^m {+} (C_1
{-} C_3)\omega^{mn}  {+}
$$
\begin{equation}
+ (C_1 {+} C_3)\sigma^{mn} + \frac13 \left(C_1 {+}C_3 \right)
\Theta \Delta^{mn} + C_2 \Theta g^{mn} \,, \label{act5a}
\end{equation}
and the corresponding term in the action functional is
$$
K^{abmn}(\nabla_a U_m) (\nabla_b U_n) =
$$
$$
=(C_1 {+} C_4)DU_k DU^k {+}
(C_1 {-} C_3)\omega_{ik}
\omega^{ik} {+}
$$
\begin{equation}
+ (C_1 {+} C_3)\sigma_{ik} \sigma^{ik} {+}   \frac13 \left(C_1 {+} 3C_2 {+}C_3 \right) \Theta^2
\,. \label{act5b}
\end{equation}

\subsection{Aether insensitive with respect to the shear and expansion of the velocity field, and solutions with pp-wave symmetry}

We consider below the background model with the so-called pp-wave symmetry. This means that we assume the following:

\noindent 1. The Lie derivative of the metric is equal to zero,
$\pounds_{\xi^l_{(\alpha)}} g_{ik} =0$, where the three Killing
vectors $\{\xi^i_{(1)}, \xi^i_{(2)}, \xi^i_{(3)} \}$ correspond to
the Abelian group of isometries ${\rm G}_3$, and $\xi^i_{(1)}$ is
the null covariant constant Killing vector, i.e.,
$g_{ik}\xi^i_{(1)}\xi^k_{(1)}=0$, and $\nabla_k \xi^i_{(1)}=0$.

\noindent 2. The vector field and pseudoscalar field inherit the
pp-wave symmetry, i.e., $\pounds_{\xi^i_{(\alpha)}} U^k =0$ and
$\pounds_{\xi^i_{(\alpha)}} \phi =0$.

\noindent Also, we consider the truncated model with  $C_2{=}0$
and $C_3{=}{-}C_1$; as it will be shown below these relationships
mean that the aether does not react on the velocity perturbations,
which are characterized by shear and expansion, and can answer on
the acceleration and vorticity only.

\noindent Then there exist solutions to the equations for the
dynamic unit vector field (\ref{0A1})-(\ref{0A309}), gravity field
(\ref{0Ein1})-(\ref{Ta}) and  pseudoscalar field (\ref{axioneq}),
which relate to vanishing stress-energy tensors of dynamic unit
vector field, $T^{({\rm U})}_{ik}{=}0$ (see(\ref{0Ein5})) and of
the pseudoscalar field, $T^{({\rm A})}_{ik}{=}0$ (see (\ref{Ta})).

\subsubsection{The proof}

Space-times with the pp-wave symmetry are well-known (see, e.g.
\cite{Exact}) As an illustration one can use the particular metric
of this class in the so-called TT-gauge, describing, a plane
gravitational wave with the first polarization \cite{MTW}
\begin{equation}
ds^2 = 2 du dv - L^2 \left(e^{2\beta} {dx^2}^2 +
e^{-2\beta} {dx^3}^2
\right) \,.
\label{PW1}
\end{equation}
Here $u$ and $v$ are the retarded and advanced times,
respectively, given in terms of the time $t$ and spatial
coordinate $x^1$ by $u{=}\frac{1}{\sqrt2}(ct{-}x^1)$,
$v{=}\frac{1}{\sqrt2}(ct{+}x^1)$, and $x^2, x^3$ are the spatial
coordinates in the plane of the front of the pp-wave. The
quantities $L(u)$ and $\beta(u)$ are  functions of the retarded
time $u$ only. The three Killing vectors in this case are known to
be of the form
\begin{equation}
\xi^i_{(1)} = \delta^i_v \,, \quad  \xi^i_{(2)} = \delta^i_2 \,, \quad \xi^i_{(3)} = \delta^i_3 \,,
\label{K1}
\end{equation}
the first of them is the null four-vector, i.e.,
$g_{ik}\xi^i_{(1)} \xi^k_{(1)} =0$; also the four-vectors are
orthogonal one to another.

From the requirement $\pounds_{\xi^i_{(\alpha)}} \phi =0$ with the
Killing vectors (\ref{K1}) one obtains that in the given
representation the function $\phi$ does not depend on $v,
x^2,x^3$, i.e., the pseudoscalar field is the function of the
retarded time $u$ only, $\phi(u)$. The master equation for the
pseudoscalar field (\ref{axioneq}) is, clearly, satisfied
identically for $\phi(u)=\Phi$. The corresponding stress-energy
tensor (\ref{Ta}) vanishes, $T^{({\rm A})}_{ik} =0$, for arbitrary
parameter $\xi$.

The requirement $\pounds_{\xi^i_{(\alpha)}} U^k =0$ with the
Killing vectors (\ref{K1}) means that the components $U^i$ can be
the functions of the retarded time only. We focus on the specific
case of this class of solutions, namely, on the case, when
\begin{equation}
U^i = \frac{1}{\sqrt2} \left(\delta^i_u + \delta^i_v \right) \,.
\label{K2}
\end{equation}
In other words, we assume that the velocity four vector has no
components orthogonal to the direction of the pp-wave propagation,
and the remaining components are constant in the chosen frame of
reference. The covariant derivative of the velocity four-vector
reduces as follows:
\begin{equation}
\Theta_i^{\ k} \equiv  \nabla_i U^k  = \frac{1}{\sqrt2} \left[\delta_i^2
\delta^k_2 \left(\frac{L^{\prime}}{L}{+}
\beta^{\prime} \right) {+} \delta_i^3 \delta^k_3
\left(\frac{L^{\prime}}{L}{-} \beta^{\prime} \right)\right]\,.
\label{GW001}
\end{equation}
Here and below the prime denotes the derivative with respect to
the retarded time $u$. Clearly, the tensor $\Theta_{ik}$ based on (\ref{GW001}) is symmetric,
i.e., $\Theta_{ik}= \Theta_{ki}$; the acceleration four-vector and the
vorticity tensor are equal to zero
\begin{equation}
DU^k = 0 \,, \quad \omega_{pq} = 0
\,.
\label{PW3bef}
\end{equation}
The expansion scalar is proportional to the derivative of the so-called background factor $L$:
\begin{equation}
\Theta = \frac{\sqrt{2}\,L^{\prime}(u)}{L} \,.
\label{GW0011}
\end{equation}
The shear tensor is also non-vanishing; it can be written as
\begin{equation}
\sigma^k_i  = \frac{\Theta}{2}  \left(\frac13
\Delta_i^k - \delta_i^1 \delta^k_1 \right)
+ \frac{\beta^{\prime}}{\sqrt2} \left(\delta_i^2 \delta^k_2
{-} \delta_i^3 \delta^k_3 \right)\,.
\label{GW002}
\end{equation}
Since the acceleration and vorticity is absent in the velocity
field, the dynamic tensor ${\cal J}^{mn}$, obtained for the case
$C_2=0$ and $C_3=-C_1$
\begin{equation}
{\cal J}^{mn} = (C_1 {+} C_4)U^m DU^n {-} C_1 U^n DU^m {+}  2C_1
\omega^{mn}   \label{act5}
\end{equation}
vanishes for arbitrary constants $C_1$ and $C_4$, i.e.,
\begin{equation}
{\cal J}^{mn}(u) = 0 \,, \quad I^n(u) = 0  \,. \label{act159}
\end{equation}
Keeping in mind (\ref{act159}) and the fact that two first terms
in (\ref{0Ein5}) disappear, when the tensor
$\Theta_{ik}=\nabla_iU_k$ is symmetric, we obtain that $T^{({\rm
U})}_{ik} =0$.

For the case, when the unit dynamic vector field is insensitive to
the shear and expansion, and the pseudoscalar field takes constant
value corresponding to one of the two minima of the axion
potential, the gravity field equations are reduced to the one
equation
\begin{equation}
\frac{L^{\prime \prime}}{L} + {\beta^{\prime}}^2 = 0 \,, \label{act53}
\end{equation}
as it should be for the model with pp-wave symmetry.
We assume that the front of the gravitational pp-wave is indicated by $u=0$, and the conditions
\begin{equation}
L(0) = 1 \,, \quad L^{\prime}(0)=0 \,, \quad \beta(0)=0  \label{act59}
\end{equation}
play the role of initial data for the metric functions.

\subsubsection{Petrov's solution}

For the illustration we consider the explicit Petrov solution \cite{Petrov}
\begin{equation}
L^2 = \cos{ku} \cdot \cosh{ku} \,, \quad 2\beta = \log{\left[\frac{\cos{ku}}{\cosh{ku}}\right]} \,.
\label{b59}
\end{equation}
(In fact, this solution possesses the symmetry related to the
group ${\rm G}_5$, which includes the mentioned group ${\rm G}_3$
as a subgroup). The corresponding interval is of the form
\begin{equation}
ds^2 = 2dudv -\cos^2{ku} \ {dx^2}^2-\cosh^2{ku} \ {dx^3}^2 \,,
\label{c59}
\end{equation}
and for this metric the Riemann tensor is covariantly constant,
i.e., $\nabla_l R^i_{\ kmn}=0$. Two non-vanishing components of
the Riemann tensor
\begin{equation}
R^2_{\ u2u} = - \left[\frac{L^{\prime \prime}}{L} +
(\beta^{\prime})^2 \right] - \left[2
\frac{L^{\prime}}{L}\beta^{\prime} + \beta^{\prime \prime} \right]
= k^2 \,, \label{qq14}
\end{equation}
\begin{equation}
R^3_{\ u3u} = - \left[\frac{L^{\prime \prime}}{L} +
(\beta^{\prime})^2 \right] + \left[2
\frac{L^{\prime}}{L}\beta^{\prime} + \beta^{\prime \prime} \right]
= - k^2 \,, \label{qq15}
\end{equation}
are constants with opposite signs, providing
$R_{uu} = 0$. The metric (\ref{c59}) is defined in the interval
$0 \leq u< \frac{\pi}{2k}$, at the end of this interval, $u=u_{{\rm \infty}}= \frac{\pi}{2k}$,
one of the metric coefficients degenerates.
The expansion scalar is now of the form
\begin{equation}
\Theta = \frac{k}{\sqrt2}\left(\tanh{ku} - \tan{ku} \right) \,, \label{q59}
\end{equation}
and the derivative $\beta^{\prime}$ (which will be necessary below) is
\begin{equation}
\beta^{\prime} = - \frac{k}{2} \left(\tanh{ku} + \tan{ku} \right) \,. \quad
\label{q57}
\end{equation}
The background pseudoscalar (axion) field is considered in this
scheme as a constant $\phi(u)= \Phi \to \phi(0)$ and will be
locally changed only in the extended model including the test
electromagnetic field (see the next Section).

\subsubsection{On the physical sense of the background solution}

The chosen particular background model is based on a specific
model of the aether, according to which the aether is insensitive
to the shear and expansion of the velocity field, but can be
active with respect to vorticity and acceleration. Such an aether
ignores the pp-wave-type perturbations, which are characterized by
the vanishing vorticity tensor and acceleration four-vector. The
corresponding pseudoscalar (axion) field is constant. When we deal
with test electrodynamic system and take into account the
so-called dynamo-optical interactions of the electromagnetic field
with the aether and the axion field, we, following the
hierarchical approach, consider the space-time and the unit vector
field to be non-perturbed; however, the local axion field is
assumed to be changed due to the interaction with the
electromagnetic field.

\section{Extended theory including the Maxwell field}

\subsection{Extended action functional}

In \cite{BL2014} the Einstein-aether theory was extended by
including all admissible terms with the Maxwell tensor $F_{ik}$.
Now we consider the particular Einstein-Maxwell-aether-axion
model, which is based on the action functional
\begin{equation}
S_{({\rm total})} = S_{({\rm B})}+ S_{({\rm EMA})}\,,
\label{EMAaction1}
\end{equation}
where the additional functional is of the form
$$
S_{({\rm EMA})} = \frac{1}{4} \int d^4 x \sqrt{{-}g} \left[C_{(0)}^{ikmn} F_{ik}F_{mn} + \phi F^{ik} F^{*}_{ik} + \right.
$$
\begin{equation}
\left. {+}  X^{pqikmn} \nabla_p U_q F_{ik}F_{mn} \right] \,.
\label{M20}
\end{equation}
Here $C_{(0)}^{ikmn}$ is the linear response tensor of a test
isotropic medium with dielectric and magnetic permittivities,
$\varepsilon$ and $\mu$, respectively; this tensor can be
represented in terms of the aether velocity $U^i$ as
$$
C_{(0)}^{ikmn}  =\frac{1}{2\mu} \left[ \left(g^{im}g^{kn} {-}
g^{in}g^{km} \right) + \right.
$$
\begin{equation}
 \left. {+} 2(\varepsilon \mu {-}1)\left(g^{i[m} U^{n]} U^k
{+} g^{k[n} U^{m]} U^i  \right)
\right] \,, \label{E6}
\end{equation}
where $p^{[i} q^{k]} \equiv \frac12 (p^iq^k{-}p^kq^i)$ denotes
anti-symmetrization. This term does not contain the covariant
derivative $\nabla_m U_n$. The tensor $X^{pqikmn}$ describes the
coupling of electromagnetic field to the non-uniformly moving
aether; is was reconstructed in \cite{BL2014} using the metric
$g_{ik}$, the covariant constant Kronecker tensors ($\delta^i_k$,
$\delta^{ik}_{ab}$ and higher order Kronecker tensors), the
Levi-Civita tensor $\epsilon^{ikab}$, and the unit vector field
$U^k$ itself. As in the previous case, $\phi$ introduces the
pseudoscalar field. The asterisk denotes the dualization, $F^{*ik}
\equiv \frac12 \epsilon^{ikmn}F_{mn}$
($\epsilon^{ikmn}=\frac{1}{\sqrt{-g}} E^{ikmn}$ with
$E^{0123}=1$).  As usual, we consider the Maxwell tensor expressed
in terms of the electromagnetic potential four-vector $A_i$
\begin{equation}
F_{ik} = \nabla_i A_k - \nabla_k A_i\,,
\label{maxwellA1}
\end{equation}
and add the equation
\begin{equation}
\nabla_k F^{*ik} = 0 \,,
\label{E9}
\end{equation}
as the direct consequence of (\ref{maxwellA1}).

\subsection{Electrodynamic equations}

The variation of the total action functional (\ref{EMAaction1}) with respect to $A_i$ gives the electrodynamic equation
\begin{equation}
\nabla_k H^{ik} = 0\,.
\label{E2}
\end{equation}
Here $H^{ik}$ is the excitation tensor linear in the Maxwell tensor
\begin{equation}
H^{ik} = C^{ikmn}F_{mn} \,,
\label{E3}
\end{equation}
where $C^{ikmn}$ is a total linear response tensor
\begin{equation}
C^{ikmn} = C_{(0)}^{ikmn} {+}  X^{pqikmn} \nabla_p U_q + \frac12 \phi \epsilon^{ikmn}\,.
\label{E5}
\end{equation}
This tensor includes three contributions: first, the contribution
from the test medium, $C_{(0)}^{ikmn}$; second, the contribution
from the dynamo-optical coupling, $X^{pqikmn} \nabla_p U_q$;
third, the contribution from the axion-photon coupling, $\frac12
\phi \epsilon^{ikmn}$.

Electrodynamics of
continuous media can be formulated in terms
of four-vectors representing
physical fields. These four-vector
fields are the electric field $E^i$, the magnetic
field ${\cal H}^i$, the electric excitation ${\cal D}^i$,
and the magnetic
excitation $B^i$ \cite{Eringen}: They are defined
in terms of $F^{ik}$ and $H^{ik}$ as,
$$
E^i = F^{ik} U_k \,, \quad B^i = F^{*ik}U_k \,,
$$
\begin{equation}
{\cal D}^i =
H^{ik} U_k \,, \quad {\cal H}^i = H^{*ik}U_k \,. \label{EHDB1}
\end{equation}
Completing the approach and inverting (\ref{EHDB1}), we find that
the tensors $F^{ik}$ and $H^{ik}$ can be written in terms of
$E^i$, $B^i$, ${\cal D}^i$, and  ${\cal H}^i$ as
$$
F^{ik} = \delta^{ik}_{mn} E^m U^n - \epsilon^{ikmn} B_m U_n \,,
$$
\begin{equation}
 H^{ik} = \delta^{ik}_{mn} {\cal D}^m U^n - \epsilon^{ikmn}
{\cal H}_m U_n \,, \label{EHDB2}
\end{equation}
where $\delta^{ik}_{mn}$ and $ \epsilon^{ikmn}$
are the generalized Kronecker delta and the Levi-Civita
tensor, respectively. Let us stress that now the four-vector $U^i$ is not a observer velocity four-vector;
it is the unit dynamic vector field associated with the velocity of the cosmic substratum.

\subsection{Extended equation for the pseudoscalar field}

The variation of the total action functional (\ref{EMAaction1})
with respect to pseudoscalar field $\phi$ gives now the extended
equation
\begin{equation}
\xi \nabla_m \nabla^m \phi {+} \left[m^2_{({\rm a})} {+}
\nu^2 (\phi^2{-}\phi^2_{*})\right]\phi {=} {-} \frac{1}{4\Psi^2_0}
F^{*}_{ik} F^{ik} \,. \label{EMA77}
\end{equation}
When the electromagnetic source in the right-hand side of this
equation is non-vanishing, the local axion field $\phi(u)$ differs
from the background constant value $\Phi$.

\subsection{On the dynamic equations for the aether velocity and for the gravity field}

We follow the hierarchical approach, and assume that when the
electrodynamic subsystem is considered to be the test one, the
corresponding additional terms appeared in the equation for the
aether velocity and for the gravity field are negligible in
comparison with the terms, which relate to the background state.
More precisely, we assume the following: first, the velocity field
is non-changed and has the same form (\ref{K2}); second, the
master equations for $U^i$ are again satisfied identically; third,
the gravity field is described again by the equation
(\ref{act53}). In other words, we solve below the equations
(\ref{E9})-(\ref{E5}) and (\ref{EMA77}) in the given background.

\subsection{Reconstruction of the tensor $X^{pqikmn}$ and independent coupling constants of the dynamo-optical interactions}

In order to represent the tensor $X^{lsikmn}$ we follow
definitions and the scheme used in \cite{BL2014}. It is well-known
that the tensor of total linear response admits the decomposition
$$
C^{ikmn} {=}  \varepsilon^{i[m}U^{n]} U^k {+} \varepsilon^{k[n} U^{m]} U^i   +
$$
$$
+\eta^{ikl}U^{[m}\nu_{l}^{\
n]} {+} \eta^{lmn}U^{[i} \nu_{l}^{\ k]} -
$$
\begin{equation}
- \frac12
\eta^{ikl}(\mu^{-1})_{ls}  \eta^{mns}
\,, \label{44}
\end{equation}
where $\varepsilon^{im}$ is the dielectric permittivity tensor,
$(\mu^{-1})_{pq}$ is the magnetic impermeability tensor,
$\nu_{p \ \cdot}^{\ m}$ is the tensor of magneto-electric
coefficients:
\begin{equation}
\varepsilon^{im} = 2 C^{ikmn} U_k U_n \,,
\label{varco199}
\end{equation}
\begin{equation}
(\mu^{-1})_{pq}  =
- \frac{1}{2} \eta_{pik}  C^{ikmn} \eta_{mnq}\,,
\label{varco1999}
\end{equation}
\begin{equation}
\nu_{p}^{\
m} = \eta_{pik} C^{ikmn} U_n =U_k C^{mkln} \eta_{lnp}\,.
\label{varco198}
\end{equation}
As usual, the tensors $\eta_{mnl}$ and $\eta^{ikl}$ are
skew-symmetric tensors orthogonal to $U^i$,
\begin{equation}
\eta_{mnl} \equiv \epsilon_{mnls} U^s \,,
\quad
\eta^{ikl} \equiv \epsilon^{ikls} U_s \,,
\label{47}
\end{equation}
and obey the following identities
\begin{equation}
- \eta^{ikp} \eta_{mnp} = \delta^{ikl}_{mns} U_l U^s = \Delta^i_m
\Delta^k_n - \Delta^i_n \Delta^k_m \,,  \label{501}
\end{equation}
\begin{equation}
-\frac{1}{2}
\eta^{ikl}  \eta_{klm} = \delta^{il}_{ms} U_l U^s =  \Delta^i_m \equiv \delta^i_m -U^iU_m
\,. \label{5019}
\end{equation}
We now decompose
explicitly the permittivity tensors $\varepsilon^{im}$,
$(\mu^{-1})_{pq}$ and $\nu^{pm}$ using the non-vanishing irreducible parts of the
covariant derivative of the velocity four-vector
$\sigma_{ik}$ and $\Theta$.
The properties
$$
\varepsilon_{ik}U^k = 0 \,, \quad {(\mu^{{-}1})}^{ik} U_k = 0 \,,
$$
\begin{equation}
\nu^{ik}U_k = 0 = \nu^{ik}U_i \label{ortho}
\end{equation}
simplify the decomposition of $\varepsilon^{im}$,
$(\mu^{-1})_{pq}$ and $\nu^{pm}$, providing the following results.
The dielectric permittivity tensor is decomposed as
\begin{equation}
\varepsilon^{ik} = \Delta^{ik} \left(\varepsilon {+} \alpha_1
\Theta \right){+}
\alpha_6 \sigma^{ik}  \,, \label{eps}
\end{equation}
where  $\Delta^{ik}=g^{ik}-U^iU^k$ is the projector,  and
$\alpha_1$ and $\alpha_{6}$ are two new independent dynamo-optical
coupling constants. The magnetic impermeability tensor is
decomposed as
\begin{equation}
{\left(\mu^{-1}\right)}^{ik} = \Delta^{ik} \left(\frac{1}{\mu}{+}
\gamma_1 \Theta \right)
{+} \gamma_6 \sigma^{ik} \,, \label{mu}
\end{equation}
where  $\gamma_1$ and $\gamma_{6}$ are the magnetic analogs of the
constants $\alpha_1$ and $\alpha_{6}$ (we use the same definitions
of the coupling constants, as in \cite{BL2014}). The
magneto-electric cross-effect pseudo-tensor is decomposed as
\begin{equation}
\nu^{pm} =  - \phi \Delta^{pm}\,,
\label{nu}
\end{equation}
i.e., the only contribution to this tensor came from the pseudoscalar field.

Finally, to reconstruct the tensor $X^{pqikmn}$ itself, we can put
(\ref{eps}) and (\ref{mu}) into the difference
$C^{ikmn}-C^{ikmn}_{(0)}-\frac 12 \phi \epsilon^{ikmn}$, using
(\ref{44}), (\ref{E6}) and (\ref{nu}), respectively. The result is
$$
X^{lsikmn} =
\frac12 \left(\alpha_1 {-} \frac13 \alpha_6 \right)
\Delta^{ls}\left(g^{ikmn}-\Delta^{ikmn} \right) +
$$
$$
+ \frac14 \alpha_6 U_p U_q \left[g^{iklp}g^{mnsq} +
g^{mnlp}g^{iksq} \right]+
$$
\begin{equation}
+ \frac12 \left(\gamma_1 {-} \frac13 \gamma_6 \right)
\Delta^{ls}\Delta^{ikmn} - \frac12 \gamma_6 \
\eta^{ik(l} \eta^{s)mn}
\,,
\label{X}
\end{equation}
i.e., this tensor contains four new coupling constants, describing
the dynamo-optical interactions. Here we used the following
auxiliary tensors
$$
g^{ikmn} \equiv g^{im}g^{kn}{-}g^{in}g^{km} \,,
$$
\begin{equation}
\Delta^{ikmn}\equiv \Delta^{im}\Delta^{kn}-\Delta^{in}\Delta^{km}\,.
\label{X99}
\end{equation}

\section{Exact solutions to the equations of extended electrodynamics in the pp-wave background}

In this Section we obtain and discuss exact solutions for the
electromagnetic and pseudoscalar fields, which possess the pp-wave
symmetry.

\subsection{Initial state}

At $u=0$ the metric functions satisfy the requirements
(\ref{act59}), and the initial value for the pseudoscalar field is
$\phi(0)$ (the detailed discussion concerning the Goursat problem
for the pseudoscalar field can be found in \cite{BWTN}). The
initial data for the electromagnetic field are indicated as
follows: $E_v(0)$, $E_2(0)$, $E_3(0)$,  $B_v(0)$, $B_2(0)$,
$B_3(0)$. As for the components $E_u$ and $B_u$, because of the
orthogonality conditions $E_i U^i {=} 0$ and $B^mU_m{=}0$, we
obtain with $U^i{=}\delta^i_0$ that for arbitrary $u$
\begin{equation}
0 =E_0=\frac{1}{\sqrt2}(E_u+E_v) \to E_u(u)=-E_v(u) \,,
\label{X6r}
\end{equation}
\begin{equation}
0 =B_0=\frac{1}{\sqrt2}(B_u+B_v) \to B_u(u)=-B_v(u) \,,
\label{Xr7}
\end{equation}
thus, $E_u(0)=-E_v(0)$, $B_u(0)=-B_v(0)$. Six initial parameters
$E_v(0)$, $E^2(0)$, $E^3(0)$,  $B_v(0)$, $B_2(0)$ and $B_3(0)$ are
linked by the requirement $E_k(0)B^k(0)=0$, which follows from the
condition that the  pseudo-invariant of the electromagnetic field
$F^{*}_{ik}F^{ik}$ is equal to zero at $u=0$, providing the
compatibility of (\ref{EMA77}) at $u=0$. In more details, this
requirement reads
\begin{equation}
2 E_v(0)B_v(0)= E^2(0)B_2(0) + E^3(0)B_3(0) \,.
\label{Xr1}
\end{equation}

\subsection{Exact solutions in terms of electric and magnetic fields}

The first step is to solve the equations (\ref{E9}). Since we
search for solutions inheriting the pp-wave symmetry and thus
depending on the retarded time only, we can follow the simple
logic way:
$$
\nabla_k F^{*ik} = 0 \to
$$
\begin{equation}
\to \frac{d}{du}(L^2 F^{*iu}) =0 \to E^{iumn}F_{mn} ={\rm const} \,.
\label{S1}
\end{equation}
Then we put $i{=}v$, $i{=}2$, $i{=}3$ and obtain three important formulas:
\begin{equation}
F_{23}(u) = F_{23}(0) \to B_v(u)= \frac{1}{L^2}B_v(0) \,,
\label{S2}
\end{equation}
$$
F_{2v}(u) = F_{2v}(0) \to
$$
\begin{equation}
\to B_3(u)= e^{-2\beta}\left[B_3(0)+E_2(u)-E_2(0) \right] \,,
\label{S3}
\end{equation}
$$
F_{3v}(u) = F_{3v}(0) \to
$$
\begin{equation}
\to B_2(u)= e^{2\beta}\left[B_2(0) - E_3(u)+ E_3(0) \right] \,.
\label{S4}
\end{equation}
These formulas allow us to replace further the components of the
magnetic field $B_v(u)$,  $B_2(u)$ and $B_3(u)$ with the electric
field components $E_v(u)$,  $E_2(u)$ and $E_3(u)$.

Then we use the same procedure for the equation (\ref{E2}) and obtain
$$
\nabla_k H^{ik} = 0 \to
$$
\begin{equation}
\to \frac{d}{du}(L^2 H^{iu}) =0 \to L^2 H^{iu}(u) ={\rm const} \,,
\label{S5}
\end{equation}
or in more details
\begin{equation}
L^2 C^{vumn}(u)F_{mn}(u) =C^{vumn}(0)F_{mn}(0) \,,
\label{S6}
\end{equation}
\begin{equation}
L^2 C^{2umn}(u)F_{mn}(u) =C^{2umn}(0)F_{mn}(0) \,,
\label{S8}
\end{equation}
\begin{equation}
L^2 C^{3umn}(u)F_{mn}(u) =C^{3umn}(0)F_{mn}(0) \,.
\label{S10}
\end{equation}
Using the representation of the linear response tensor $C^{ikmn}$
(\ref{E5})  with $C^{ikmn}_{(0)}$ from (\ref{E6}) and $X^{pqikmn}$
from (\ref{X}), as well as, the formulas (\ref{S2})-(\ref{S4}), we
obtain three components of the electric field. First, we display
the longitudinal (with respect to the pp-wave front) component of
the electric field:
\begin{equation}
\Delta_{(v)}(u) \cdot E_v(u)= \varepsilon E_v(0) + B_v(0)[\phi(u)-\phi(0)] \,,
\label{S7}
\end{equation}
where
\begin{equation}
\Delta_{(v)}(u) \equiv L^2\left[\varepsilon+ \Theta\left(\alpha_1-\frac13 \alpha_6 \right)\right] \,.
\label{S7a}
\end{equation}
Second, we display the transversal component $E^2(u)$ in the form
$$
\Delta_{(2)}(u) \cdot E^2(u)= E^2(0)\left(\varepsilon {-} \frac{1}{\mu} \right) + [E^2(0){+} B_3(0)] \times
$$
$$
\times \left\{\frac{1}{\mu}\left(1{-}e^{{-}2\beta} \right){-}
e^{{-}2\beta}\left[\Theta\left(\gamma_1{+}\frac16 \gamma_6 \right)
{-} \frac{\beta^{\prime}}{\sqrt2} \gamma_6  \right] \right\} {-}
$$
\begin{equation}
{-}[E^3(0){-} B_2(0)][\phi(u){-}\phi(0)]  \,,
\label{S9}
\end{equation}
where
$$
\Delta_{(2)}(u) \equiv L^2\left\{\left(\varepsilon {-}
\frac{1}{\mu} \right) + \frac{1}{\sqrt2} \beta^{\prime}
(\alpha_6{+} \gamma_6) + \right.
$$
\begin{equation}
\left. + \Theta \left[(\alpha_1{-}\gamma_1)+\frac16(\alpha_6{-}\gamma_6) \right] \right\} \,.
\label{S9a}
\end{equation}
Finally, we obtain for the transversal component $E^3(u)$
$$
\Delta_{(3)}(u) \cdot E^3(u)= E^3(0)\left(\varepsilon {-} \frac{1}{\mu} \right) + [E^3(0){-} B_2(0)] \times
$$
$$
\times \left\{\frac{1}{\mu}\left(1-e^{2\beta} \right)- e^{2\beta}
\left[\Theta\left(\gamma_1+\frac16 \gamma_6 \right) +
\frac{\beta^{\prime}}{\sqrt2} \gamma_6 \right] \right\} +
$$
\begin{equation}
+ [E^2(0){+} B_3(0)][\phi(u){-}\phi(0)] \,,
\label{S11}
\end{equation}
where the term
$$
\Delta_{(3)}(u) \equiv L^2\left\{\left(\varepsilon {-}
\frac{1}{\mu} \right) - \frac{1}{\sqrt2}
\beta^{\prime}(\alpha_6{+} \gamma_6) + \right.
$$
\begin{equation}
\left. + \Theta \left[(\alpha_1{-}\gamma_1)+ \frac16(\alpha_6{-}\gamma_6) \right] \right\}
\label{S11a}
\end{equation}
can be obtained from $\Delta_{(2)}(u)$ by the formal replacement
$\beta \to - \beta$. Thus, the electric and magnetic fields are
expressed through the metric functions $L(u)$, $\beta(u)$ and
their derivatives, $L^{\prime}(u)$, $\beta^{\prime}(u)$, and
through the variation of the pseudoscalar field
$[\phi(u){-}\phi(0)]$.

\subsection{Equation for the pseudoscalar (axion) field}

When $\phi$ depends on the retarded time only, the equation for the pseudoscalar field (\ref{EMA77}) is reduced to
\begin{equation}
F^{*}_{ik}F^{ik} = -4\Psi^2_0 \nu^2 \phi \left[\phi^2(u)-\Phi^2 \right] \,,
\label{axion44}
\end{equation}
which is, in fact, the implicit equation for the function
$\phi(u)$. Using (\ref{S2})-(\ref{S4}) this equation can be
simplified as
$$
\Psi^2_0 \nu^2 \phi \left[\phi^2(u)-\Phi^2 \right] = {-} E^2(u) e^{2\beta} \left[B_2(0){-}E^3(0)\right] -
$$
\begin{equation}
{-}
E^3(u) e^{{-}2\beta} \left[B_3(0){+}E^2(0)\right] + \frac{2}{L^2} E_v(u)B_v(0)\,.
\label{S13}
\end{equation}
Since $E^2(u)$ and $E^3(u)$ are linear with respect to $\phi(u)$,
we deal with cubic equations for the axion field. We will return
to this equation in the next Section.

\section{Anomalies induced by pp-waves}

\subsection{Longitudinal magnetic and electric fields}

In order to simplify the physical analysis of the formulas
presented in general form (\ref{S7})-(\ref{S11a}), we consider
separately the longitudinal and transversal configurations of the
electromagnetic fields. When $E^2(0)=E^3(0)=0$ and
$B_2(0)=B_3(0)=0$, only the longitudinal electric and magnetic
fields exist in the electrodynamic system at $u>0$. The condition
(\ref{Xr1}) takes now the form $E_v(0)B_v(0)=0$, thus, it is
satisfied in two cases only.

\subsubsection{$E_v(0) \neq 0$,  $B_v(0)=0$}

In this case we obtain  from (\ref{S7}), (\ref{S7a}) and (\ref{S2}), that
\begin{equation}
E_v(u)= \frac{\varepsilon E_v(0)}{L^2\left[\varepsilon+
\Theta\left(\alpha_1-\frac13 \alpha_6 \right) \right]} \,,\quad
B_v(u)=0 \,. \label{S14}
\end{equation}
The equation (\ref{S13}) yields in this case that $\phi(u)=\Phi$, i.e.,
the background pseudoscalar field is not disturbed.

The solution for the longitudinal electric field can grow
anomalously, when the denominator in (\ref{S14}) tends to zero. As
an illustration we consider the example of the Petrov solution and
obtain
\begin{equation}
E_v(u)= \frac{E_v(0)}{\Delta_{||}(u)\cosh{ku}}
\,,
\label{S19}
\end{equation}
where
$$
\Delta_{||}(u) \equiv \cos{ku}\left[1+\frac{k}{\sqrt2 \varepsilon}\left(\alpha_1-\frac13 \alpha_6 \right) \tanh{ku} \right] -
$$
\begin{equation}
-
\frac{k}{\sqrt2 \varepsilon}\left(\alpha_1-\frac13 \alpha_6 \right) \sin{ku}  \,.
\label{S20}
\end{equation}
Clearly,  $\Delta_{||}(0){=}1>0$ and $\Delta_{||}(\frac{\pi}{2k}){=}
\frac{k}{\sqrt2 \varepsilon}\left(\frac13 \alpha_6 {-}\alpha_1\right)$. This means that, when $\alpha_1>\frac13 \alpha_6$
the quantity $\Delta_{||}(\frac{\pi}{2k})$ is negative, thus, there
is a moment $u_{**}$, for which $\Delta_{||}(u_{**})=0$, so that
$E_v(u_{**})=\infty$  i.e., the anomaly exists. When
$\alpha_1<\frac13 \alpha_6$ the longitudinal electric field
reaches asymptotically the value
\begin{equation}
E_v\left(\frac{\pi}{2k}\right)= \frac{\sqrt2 \varepsilon E_v(0)}{k \left(\frac13 \alpha_6 -\alpha_1\right) \cosh{\frac{\pi}{2}}}
\,,
\label{S21}
\end{equation}
which is much bigger than $E_v(0)$ for small parameter $k
\left(\frac13 \alpha_6 -\alpha_1 \right)$. Let us stress that in
this case the longitudinal electric field is finite, when the
metric is degenerated.

\subsubsection{$E_v(0) = 0$,  $B_v(0) \neq 0$}

Now the longitudinal magnetic field is deformed as $B_v(u)=\frac{B_v(0)}{L^2}$,  and the longitudinal electric field
\begin{equation}
E_v(u)= \frac{B_v(0)[\phi(u)-\phi(0)]}{L^2\left[\varepsilon+
\Theta \left(\alpha_1-\frac13 \alpha_6 \right)\right]}
\label{S789}
\end{equation}
is, formally speaking, generated by the axion field. The equation
for the axion field (\ref{S13}) transforms into the following
cubic equation:
\begin{equation}
(\phi{-}\Phi)\left\{\phi^2{+}\Phi \phi - \frac{2B^2_v(0)}{L^4
\Psi^2_0 \nu^2 \left[\varepsilon{+}  \Theta\left(\alpha_1{-}\frac13
\alpha_6 \right)\right]}\right\}=0\,. \label{S1345}
\end{equation}
Clearly, only one solution to this equation, $\phi(u)=\Phi$
satisfies the initial condition in general case, and we have to
state that the axion field is not modified, and the longitudinal
electric field can not be induced. But there is a special
(critical) case, when the initial longitudinal magnetic field
takes the value
\begin{equation}
B_v(0)= B_{\rm critical} = \Phi \Psi_0 \nu \sqrt{\varepsilon} \,,
\label{S1395}
\end{equation}
and the corresponding equation for the axion field
\begin{equation}
(\phi{-}\Phi)\left\{\phi^2{+}\Phi \phi - \frac{2\Phi^2 \varepsilon}{L^4 \left[\varepsilon{+} \Theta\left(\alpha_1{-}\frac13 \alpha_6 \right)\right]}\right\}=0
\label{S1399}
\end{equation}
in addition to the constant solution $\phi(u) \equiv \Phi$ admits the solution
\begin{equation}
\phi(u) = \frac12 \Phi \left\{\sqrt{1{+} \frac{8 \varepsilon}{L^4 \left[\varepsilon{+} \Theta\left(\alpha_1{-}\frac13 \alpha_6 \right)\right]}} {-}1 \right\}   \,,
\label{zS1399}
\end{equation}
which has the initial value $\phi(0)=\Phi$, and is of an anomalous type, when $\alpha_1>\frac13 \alpha_6 $.
We deal now with a bifurcation of the solutions for the axion field.

\subsection{The model with initially transversal pure magnetic field}

In this Subsection we consider the model with initial data $B_v(0)=0$, $E_v(0)=0$,  $E^2(0)=0$, $E^3(0)=0$.
Clearly, in this case the longitudinal electric and magnetic field are absent for arbitrary $u>0$.

\subsubsection{Exact solutions for the transversal electric and magnetic fields}

Transversal magnetic field interacting with gravitational and pseudoscalar fields generates the transversal electric field with the following components:
$$
E^2(u)= \frac{B_3(0)}{\Delta_{(2)}}\left\{\frac{1}{\mu}\left(1{-}e^{{-}2\beta} \right){-} e^{{-}2\beta}\left[ \Theta\left(\gamma_1{+}\frac16 \gamma_6 \right) {-}  \right.\right.
$$
\begin{equation}
\left.\left. -\frac{\beta^{\prime}}{\sqrt2} \gamma_6 \right] \right\}  +\frac{B_2(0)}{\Delta_{(2)}}[\phi(u){-}\phi(0)]  \,,
\label{S23}
\end{equation}
$$
E^3(u)= -\frac{B_2(0)}{\Delta_{(3)}}\left\{\frac{1}{\mu}\left(1-e^{2\beta} \right)-e^{2\beta}\left[ \Theta\left(\gamma_1+\frac16 \gamma_6 \right) + \right. \right.
$$
\begin{equation}
\left. \left. + \frac{\beta^{\prime}}{\sqrt2} \gamma_6 \right] \right\} + \frac{B_3(0)}{\Delta_{(3)}}[\phi(u){-}\phi(0)]     \,.
\label{S24}
\end{equation}
Clearly, $E^2(u \to 0)=0$ and $E^3(u \to 0)=0$, covering the initial data, i.e., the electric field is, indeed, generated by the gravitational wave in the axionic environment.
With these formulas the magnetic field components $B_2(u)$ and $B_3(u)$ can be obtained as follows:
$$
B_2(u)= e^{2\beta}B_2(0) + L^2 E^3(u) =
$$
$$
= \frac{L^2B_2(0)}{\Delta_{(3)}}\left\{\left(\varepsilon e^{2\beta} -\frac{1}{\mu}\right) + e^{2\beta}\left[ \Theta\left(\alpha_1+\frac16 \alpha_6 \right) - \right. \right.
$$
\begin{equation}
\left. \left. -\frac{\beta^{\prime}}{\sqrt2} \alpha_6 \right] \right\} - \frac{L^2 B_3(0)}{\Delta_{(3)}}[\phi(u){-}\phi(0)] \,,
\label{11S3}
\end{equation}
$$
B_3(u)= e^{-2\beta} B_3(0)- L^2 E^2(u) =
$$
$$
= \frac{L^2B_3(0)}{\Delta_{(2)}}\left\{\left(\varepsilon e^{-2\beta} -\frac{1}{\mu}\right) + e^{-2\beta}\left[ \Theta\left(\alpha_1+\frac16 \alpha_6 \right) + \right. \right.
$$
\begin{equation}
\left. \left.+\frac{\beta^{\prime}}{\sqrt2} \alpha_6 \right] \right\} - \frac{L^2B_2(0)}{\Delta_{(2)}}[\phi(u){-}\phi(0)]\,.
\label{12S3}
\end{equation}
Formally speaking, the denominators in the formulas (\ref{S23})-(\ref{12S3}) can take zero values providing the anomalies
in the responses of the electromagnetic and pseudoscalar fields. In order  to clarify these possibilities in detail,
we consider below an explicit example as an illustration of this anomaly.

\subsubsection{Illustration of anomalous behavior on the base of the explicit Petrov's solution}

We consider the formula for $E^2(u)$ only, since the results for $E^3$ can be obtained by the formal replacement
$\beta \to - \beta$. Let us focus on the function $\Delta_{(2)}(u)$ given by (\ref{S9a}) and appeared in
the denominator of the function $E^2(u)$ (\ref{S23}), and let us transform it for the case, when the metric functions are given by the Petrov formulas (\ref{b59}):
$$
\Delta_{(2)}(u) = \cosh{ku}\left\{\cos{ku}\left[ \left(\varepsilon {-} \frac{1}{\mu} \right) + \right. \right.
$$
$$
\left. \left. + \frac{k}{\sqrt2}  \tanh{ku} \left( (\alpha_1{-}\gamma_1)- \frac13(\alpha_6{+}2\gamma_6) \right)\right] - \right.
$$
\begin{equation}
\left. - \frac{k}{\sqrt2}\sin{ku} \left[ (\alpha_1{-}\gamma_1)+ \frac13(2\alpha_6{+}\gamma_6) \right] \right\} \,.
\label{S31}
\end{equation}
At $u=0$ and $u=\frac{\pi}{2k}$ this function takes the values, respectively,
$$
\Delta_{(2)}(0)=\left(\varepsilon {-} \frac{1}{\mu} \right) \,,
$$
\begin{equation}
\Delta_{(2)}\left(\frac{\pi}{2k}\right) = - \frac{k}{\sqrt2} \cosh{\frac{\pi}{2}}\left[ (\alpha_1{-}\gamma_1)+ \frac13(2\alpha_6{+}\gamma_6) \right]  \,.
\label{S55}
\end{equation}
For the standard medium the refraction index exceeds one, i.e., $\varepsilon > \frac{1}{\mu}$, thus, $\Delta_{(2)}(0)$ is positive.
When $\Delta_{(2)}\left(\frac{\pi}{2k}\right)<0$, i.e.,
$\frac13(2\alpha_6{+}\gamma_6) > \gamma_1 {-} \alpha_1$, we deal again with the anomaly in the electric field at the moment $u_{**}$,
in which
$\Delta_{(2)}\left(u_{**}\right)=0$. When $\frac13(2\alpha_6{+}\gamma_6) < \gamma_1 {-} \alpha_1$, the electric component $E^2$
tends asymptotically to the finite value $E^2\left(\frac{\pi}{2k}\right)$. Let us mention, that the component $E^3$ remains finite,
when $E^2=\infty$, and vice-versa, $E^2$ is  finite, when $E^3=\infty$.

The behavior of the function $\Delta_{(2)}(u)$ is more illustrative, if the unknown coupling constants are linked,
e.g., by the relationship $3(\alpha_1{-}\gamma_1){=}\alpha_6{+}2\gamma_6$. In this case the solution $u_{**}$
to the equation $\Delta_{(2)}(u_{**}){=}0$ can be found explicitly as
\begin{equation}
u_{**} = \frac{1}{k} {\rm arcctg}\left[\frac{k (\alpha_6{+}\gamma_6)}{\sqrt2 \left(\varepsilon-\frac{1}{\mu} \right)}\right] \,.
\label{Sxx}
\end{equation}

\subsubsection{Exact solutions for the pseudoscalar (axion) field}

The pseudoscalar field can be found now from the equation
$$
\Psi^2_0 \nu^2 \phi \left[\phi^2(u)-\Phi^2 \right] =
$$
\begin{equation}
- E^2(u) e^{2\beta} B_2(0) -
E^3(u) e^{-2\beta} B_3(0) \,.
\label{aS13}
\end{equation}
Using the definitions
\begin{equation}
B_2(0) = B_{\bot} \cos{\theta} \,, \quad B_3(0) = B_{\bot} \sin{\theta} \,,
\label{S251}
\end{equation}
we reduce this equation to the standard form of the cubic equation
\begin{equation}
\phi^3 + {\cal P} \phi + {\cal Q} =0 \,,
\label{cub1}
\end{equation}
where the coefficients ${\cal P}$ and ${\cal Q}$ have the form
$$
{\cal P} = - \Phi^2 - \frac{B^2_{\bot}L^2}{\Psi^2_0 \nu^2 \Delta_{2} \Delta_{3}}
\left\{\frac{\beta^{\prime}}{\sqrt2} (\alpha_6{+} \gamma_6){\cal H}_1(u,\theta)- \right.
$$
\begin{equation}
\left. - {\cal G}(u) {\cal H}_2(u,\theta)\right\} \,,
\label{cub2}
\end{equation}
$$
{\cal Q} =
\frac{L^2 B^2_{\bot}}{\mu \Psi^2_0 \nu^2 \Delta_2 \Delta_3} \left\{ \mu \Phi \frac{\beta^{\prime}}{\sqrt2} (\alpha_6{+} \gamma_6){\cal H}_1(u,\theta) {+}  \right.
$$
$$
\left. +
\frac{\beta^{\prime}}{\sqrt2} (\alpha_6{+} \gamma_6)
\left[1{-}\cosh{2\beta} {+} \mu \Theta \left(\gamma_1{+} \frac16 \gamma_6 \right)  \right]\sin{2\theta}{+}
\right.
$$
\begin{equation}
\left.  {+} \left[\left(\sinh{2\beta} {+} \frac{1}{\sqrt2} \mu \gamma_6 \beta^{\prime} \right)\sin{2\theta}
{-} \mu \Phi {\cal H}_2(u,\theta) \right]{\cal G}(u)
 \right\}\,.
\label{cub3}
\end{equation}
Here the auxiliary functions ${\cal G}(u)$, ${\cal H}_1(u,\theta)$ and ${\cal H}_2(u,\theta)$ are introduced as follows
$$
{\cal G}(u) =\left(\varepsilon {-} \frac{1}{\mu} \right) + \Theta \left[(\alpha_1{-}\gamma_1)+ \frac16(\alpha_6{-}\gamma_6) \right] \,,
$$
$$
{\cal H}_1(u,\theta) = \cosh{2\beta}{+} \cos{2\theta}\sinh{2\beta} \,,
$$
\begin{equation}
{\cal H}_2(u,\theta) = \sinh{2\beta}{+} \cos{2\theta}\cosh{2\beta} \,.
\label{S29}
\end{equation}
At $u=0$ the quantities ${\cal P}$ and ${\cal Q}$ take the values
$$
{\cal P}(0) = -\Phi^2 {+} {\cal A} \,, \quad {\cal Q}(0) = {-} {\cal A} \Phi \,,
$$
\begin{equation}
{\cal A} = \frac{\mu B^2_{\bot} \cos{2\theta}}{\Psi^2_0 \nu^2 (\varepsilon \mu {-}1)}\,.
\label{Sw1}
\end{equation}
The corresponding cubic equation reduces at $u=0$ to
\begin{equation}
[\phi(0)-\Phi][\phi^2(0) + \phi(0) \Phi + {\cal A}]=0 \,,
\label{Sw2}
\end{equation}
providing one of solutions to be $\phi(0)=\Phi$, as we need for compatibility of the model.
The discriminant ${\cal D}$ of the cubic equation (\ref{cub1}),
${\cal D}= - (4 {\cal P}^3 + 27 {\cal Q}^2)$, calculated with ${\cal P}$ and ${\cal Q}$,
given by (\ref{cub2}) and (\ref{cub3}), respectively, regulates the properties of the roots
of the equation (\ref{cub1}). It is well-known that there are three real roots, when ${\cal D}>0$;
two real roots from three coincide, when ${\cal D}=0$; there is only one real root, when ${\cal D}<0$.
The corresponding Cardano formulas are well documented. Since the discriminant ${\cal D}$ is a sophisticated
function of retarder time and coupling parameters, we restrict ourselves by one simple case, illustrating our findings.
Let the free parameters be chosen as follows:
\begin{equation}
\alpha_1=\alpha_6=\gamma_6=0 \,, \quad \theta = \frac{\pi}{4} \,, \quad \mu = \frac{1}{\Phi} \,.
\label{Sad}
\end{equation}
Then, one obtains that
\begin{equation}
{\cal Q} = 0 \,, \quad {\cal P} = -\Phi^2 +
\frac{B^2_{\bot} \sinh{2\beta}}{\Psi^2_0 \nu^2 L^2 \left[(\varepsilon {-} \Phi){-} \gamma_1 \Theta\right]}  \,,
\label{Sad2}
\end{equation}
and the solutions to (\ref{cub1}) are
$$
\phi_1(u) = 0 \,,
$$
\begin{equation}
\phi_{2,3}(u) = \pm \sqrt{\Phi^2 -
\frac{B^2_{\bot} \sinh{2\beta}}{\Psi^2_0 \nu^2 L^2 \left[(\varepsilon {-} \Phi){-} \gamma_1 \Theta \right] }} \,.
\label{Sad3}
\end{equation}
When $u=0$ the set of these three functions covers the set of background solutions $\phi \to \{0,\pm \Phi\}$.
When we deal with the Petrov metric (\ref{c59}), the distinguished solution with initial value $+\Phi$ behaves as follows:
\begin{equation}
\phi_{2}(u) =  \sqrt{\Phi^2 +
\frac{B^2_{\bot} (\cosh^2{ku}-\cos^2{ku})}{2\Psi^2_0 \nu^2 {\cal F}(u)  \cos{ku}\cosh{ku}}}  \,.
\label{Sad4}
\end{equation}
The auxiliary function ${\cal F}(u)$ given by
$$
{\cal F}(u) = (\varepsilon {-} \Phi)\cos{ku}\cosh{ku} {-}
$$
\begin{equation}
-\frac{\gamma_1 k}{\sqrt2} (\sinh{ku} \cos{ku}-\cosh{ku}\sin{ku}) \,,
\label{Sad5}
\end{equation}
takes the following values at $u=0$ and $u=\frac{\pi}{2k}$, respectively:
\begin{equation}
{\cal F}(0) = (\varepsilon {-} \Phi) \,, \quad
{\cal F}\left(\frac{\pi}{2k} \right) = \frac{\gamma_1 k}{\sqrt2} \cosh{\frac{\pi}{2}} \,.
\label{Sad6}
\end{equation}
This means that, when $\gamma_1<0$, there exists a point, $u=u_{*}$ in which ${\cal F}(u_{*})=0$,
providing the value $\phi_{2}(u_{*})$ to be infinite.  When $\gamma_1>0$ the singularity appears
at the end of admissible interval, i.e., at $u=\frac{\pi}{2k}$.

\section{Discussion}

The main result of the presented work is the prediction that the interaction between a unit dynamic
vector field, attributed to the macroscopic velocity of a cosmic substratum (aether, vacuum, dark matter, etc.),
on the one hand, and an electrodynamic system, on the other hand, can provoke an anomalous electromagnetic response
on the action of a pp-wave gravitational field. In fact, we deal with the fifth model, which predicts the anomalous
behavior of the electromagnetic response on the impact of the plane gravitational wave. In \cite{BV} exact solutions
were obtained, which described the anomalous response of initially static magnetic field in a simple dielectric medium
at rest. In that work the amplitude of the signal-response happened to be proportional to the factor $(n^2 -1)^{-1}$,
so, the anomaly was associated with the singularity, which occurs in the vicinity of unit refraction index $n \equiv \varepsilon \mu=1$.
Such anomaly can be indicated as the static one. According to results obtained in \cite{BWTN}, static anomaly in the vicinity
of $n=1$ exists also for the electrodynamic system surrounded by the pseudoscalar (axion) field. In \cite{BL} it was shown that
in addition to the static anomaly there exists a dynamic anomaly induced by non-minimal coupling of electromagnetic and gravitational
fields. This anomaly appears at some moment of the retarded time $u_{*}$, at which the denominator in the expression for the amplitude
of the induced electric field takes zero value. Analogous behavior is typical for moving electrodynamically active media, when the medium
motion is non-uniform (see \cite{BAlpin}); this dynamic anomaly can be indicated as the dynamo-optical one.
In this context, the results obtained in the present work generalize results, which concern both static and dynamic anomalies
in the moving electrodynamic system in a pseudoscalar (axion) environment; this generalization includes couplings of the electromagnetic
field to a unit dynamic vector field associated with the velocity of a distinguished cosmic substratum, e.g., aether, vacuum, etc.

In order to illustrate new results let us focus on the formula (\ref{S23}) supplemented by (\ref{b59}), (\ref{S31}), (\ref{Sad4}),
and (\ref{Sad5}). The transversal component $E^2(u)$ of the electric field tends to infinity in two cases: first,
when $\Delta_{(2)}(u) \to 0$ (see (\ref{S31})); second, when $\phi_2(u) \to \infty$ (see (\ref{Sad4}), and (\ref{Sad5})).
When the coupling constants $\alpha_1$, $\alpha_6$, $\gamma_1$ and $\gamma_6$ vanish, the quantity $\Delta_{(2)}$
is proportional to $\epsilon \mu {-}1$, thus, this limit covers the results of \cite{BV,BL,BWTN} concerning the static anomalies in the vicinity of $n{=}1$.

When the mentioned coupling constants are non-vanishing, we deal with dynamic anomalies of two types. The dynamic anomaly of the first type
is displayed at the moment $u=u_{**}$, for which $\Delta_{(2)}(u_{**})=0$ (see also the  explicit formula (\ref{Sxx}) as an  illustration).
 The amplification of the electromagnetic response is produced by the coupling of the electromagnetic field to the unit dynamic vector field,
 which plays the role of an energy reservoir. The dynamic anomaly of the second type appears, when $\phi(u_{+})=\infty$, where $u_{+}$ is
 the root of the equation ${\cal F}(u_{+})=0$ (see (\ref{Sad5})). Now the amplification of the electric field is mediated by the anomalous
 growth of the pseudoscalar (axion) field, which, in its turn, is affected by the axion-photon coupling. Again, the unit dynamic vector field
 plays the role of an energy reservoir for the amplification of electromagnetic and pseudoscalar (axion) fields. As in the models without
 pseudoscalar field mentioned above, the role of the gravitational pp-wave is provocative: it manages the process of energy redistribution
 between unite dynamic vector field (the reservoir) and photons (the test subsystem) in the environment of axions (the mediator).

\vspace{5mm}
\noindent
{\bf Acknowledgments}

\noindent Authors thank financial support from the Program of Competitive
Growth of KFU Project No.~0615/06.15.02302.034 and from the Russian
Foundation for Basic Research (Grants RFBR No.~14-02-00598 and No.~15-52-05045).
ABB is grateful to Professor J.P.S. Lemos (CENTRA, IST, Lisbon) for fruitful discussions and hospitality, and
acknowledges financial support provided under the European Union's
FP7 ERC Starting Grant ``The dynamics of black holes:~testing the
limits of Einstein's theory" grant agreement No.~DyBHo-256667.

\end{document}